\documentclass[runningheads]{llncs}
\usepackage[T1]{fontenc}
\usepackage[dvipsnames]{xcolor}
\usepackage{graphicx}  % For rotation
\usepackage{pdflscape} % For rotating the entire page
\usepackage{array}     % For column width customization
\usepackage{adjustbox} % For fine-tuned table adjustment
\usepackage{svg} % for including the ORCID icon
\svgpath{{./}}  % If orcid.svg is in the same folder as .tex
\usepackage{lscape}
\usepackage{graphicx}
\usepackage{hyperref}
\usepackage{tcolorbox}
\usepackage{lipsum}
\usepackage{enumitem}
\usepackage{tikz}
\usepackage{amsmath,amssymb,amsfonts}
\usepackage{algorithmic}
\usepackage{graphicx}
\usepackage{listings}
\usepackage{textcomp}
\usepackage{xcolor}
\usepackage{comment}
\usepackage{hyperref}
\usepackage{balance}
\usepackage{lscape}
\usepackage{soul}
\usepackage{enumitem}
\usepackage{marvosym} % provides \Letter symbol

\usepackage{graphicx}
\usepackage{todonotes}

\newcommand{\orcid}[1]{\href{https://orcid.org/#1}{\includegraphics[height=1.5ex]{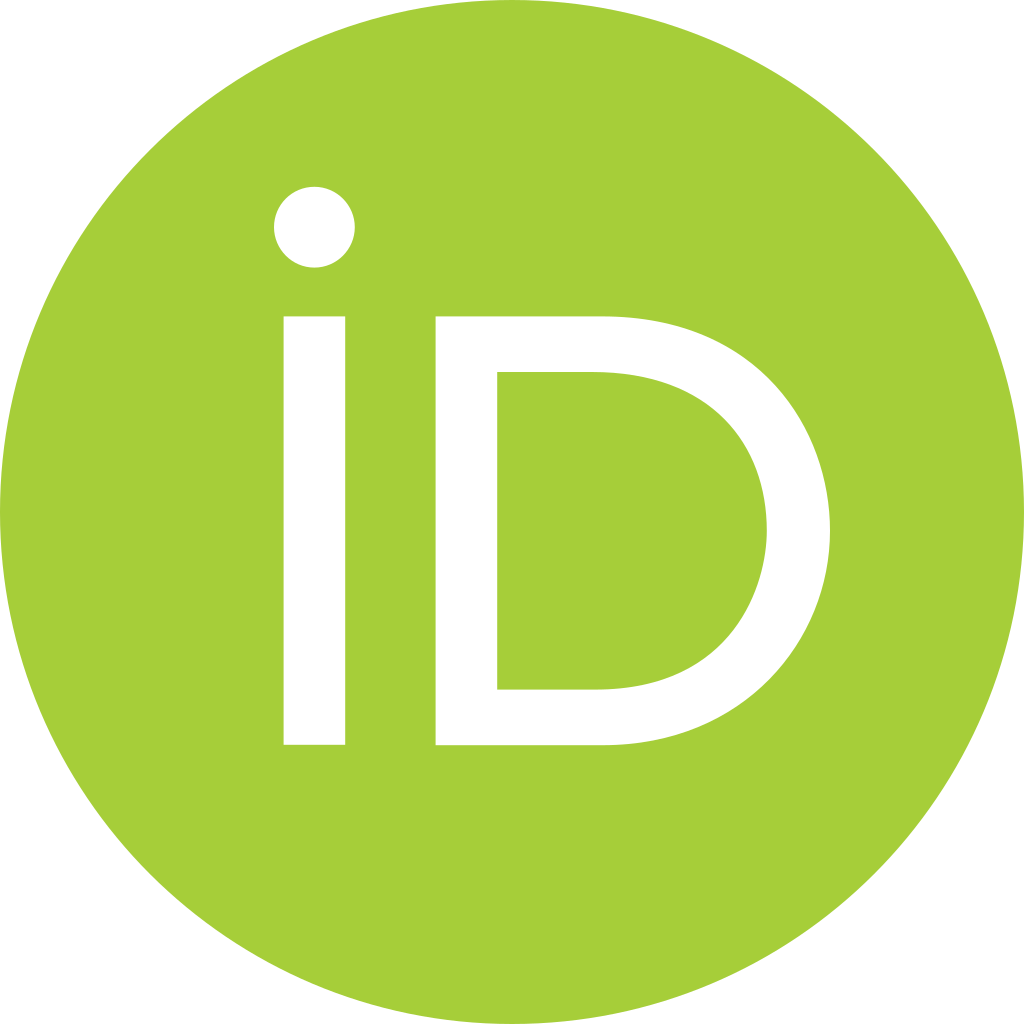}}}

\def\BibTeX{{\rm B\kern-.05em{\sc i\kern-.025em b}\kern-.08em
    T\kern-.1667em\lower.7ex\hbox{E}\kern-.125emX}}

\begin{document}

\begin{tcolorbox}[colback=yellow!10!white, colframe=yellow!50!black]
\centering \textbf{This is a preprint version. The final version will appear in the proceedings of SafeComp 2025.}
\end{tcolorbox}

\title{Large Language Models in Code Co-generation for Safe Autonomous Vehicles}
\titlerunning{LLMs in Code Co-generation for Safe Autonomous Vehicles}

\author{Ali Nouri\inst{1,3}\textsuperscript{(\Letter)}\orcid{0000-0002-9634-6094} \and
Beatriz Cabrero-Daniel\inst{2, 3} \and 
Zhennan Fei\inst{1,3} \and \\
Krishna Ronanki\inst{2, 3} \and
Håkan Sivencrona\inst{1} \and 
Christian Berger \inst{2, 3}}
\authorrunning{A. Nouri et al.}
\institute{Volvo Cars, Gothenburg, Sweden \\
\email{\{ali.nouri, zhennan.fei, hakan.sivencrona\}@volvocars.com}\\
\and
University of Gothenburg, Sweden\\
\email{\{beatriz.cabrero-daniel, ronanki, christian.berger\}@gu.se}\\
 \and
Chalmers University of Technology, Gothenburg, Sweden\\
\email{}}
\maketitle

\begin{abstract}
Software engineers in various industrial domains are already using Large Language Models (LLMs) to accelerate the process of implementing parts of software systems. When considering its potential use for ADAS or AD systems in the automotive context, there is a need to systematically assess this new setup: LLMs entail a well-documented set of risks for safety-related systems' development due to their stochastic nature.
To reduce the effort for code reviewers to evaluate LLM-generated code, we propose an evaluation pipeline to conduct sanity-checks on the generated code. We compare the performance of six state-of-the-art LLMs (CodeLlama, CodeGemma, DeepSeek-r1, DeepSeek-Coders, Mistral, and GPT-4) on four safety-related programming tasks.
Additionally, we qualitatively analyse the most frequent faults generated by these LLMs, creating a failure-mode catalogue to support human reviewers. Finally, the limitations and capabilities of LLMs in code generation, and the use of the proposed pipeline in the existing process, are discussed.

\end{abstract}

\section{Introduction}
Function realisations and improvement in software-defined vehicles require continuous software updates; hence, rapid, efficient, and continuous software development is crucial to maintaining competitiveness and user satisfaction. LLMs can be seen as a potential element in the software development pipeline, as their capability in code generation has been demonstrated~\cite{Liu2023EmpiricalStudy}. However, the limitations and capabilities of LLMs are under-explored, as they are examined primarily for simple coding tasks and less for complex and novel tasks that require creativity.

Generating code with LLMs might require multiple tries, given the LLMs' stochastic behaviour~\cite{Bender2021LLMsTooBig} and the complexity of the task. Moreover, as an LLM does not possess a proper understanding of the real world, it might fail to propose an appropriate strategy in the generated code, that may not be easily detected by a human reviewer. Hence, it would be beneficial to integrate a preliminary assessment of the generated code against unintended behaviours using key safety indicators before submitting it for review. Having this preliminary feedback on the code also helps with the automatic selection of the best model for each task and supports model improvement in the reinforcement learning process.

To assess the viability, limitations, and capabilities of LLMs, and to propose appropriate adaptations in the development of safety-related software, the following research questions have been formulated:

\begin{enumerate}[leftmargin=1.05cm]
    \item[\textbf{RQ1}:] How do state-of-the-art LLM models perform in code generation tasks for automotive functions? \textit{Quantitative evaluations of generations in Sec.~\ref{sec:results}}
    \item[\textbf{RQ2}:] What are the key limitations and risks of using LLM-generated code in safety-related applications? \textit{Qualitative analysis of LLM failures in Sec.~\ref{sec:results}}
    \item[\textbf{RQ3:}] What adaptations to existing software engineering processes are necessary to safely integrate LLM-generated code into automotive software development workflows? \textit{Proposal of LLM-augmented review, verification, and validation processes for safety-related driving functions in Sec.~\ref{sec:VRandCR}.}
\end{enumerate}
In this paper, we propose an approach to ``Generate fast, Eliminate fast'' by integrating a rapid checker within the code generation process to reduce the time a code reviewer is spending to check useless code that might not even compile. Additionally, it ensures that the unsafe software is not delivered to further stages in the V\&V process, where more rigorous and time-consuming tests, like in Hardware-in-the-Loop (HIL)-environments, would take place. 
The proposed Software-in-the-Loop (SIL) code co-generation environment combines LLM-based code generation for automotive software with an automatic assessment in a virtual test environment. The results from the simulation are used in automated evaluation and ranking of the codes and sent to the user for reviewing the code.
In order to establish the true capabilities and limitations of the used LLMs, fairly evaluate them, and indirectly validate the results, we propose a specific test design strategy. First, we avoid to give extra help such as conversations or skeleton of the code to the LLM to identify their practical capability. Second, we design tests to avoid \textit{benchmark leakage}~\cite{zhou2023don}, that implies the risk of testing LLMs' memory (since common driving functions and their benchmarks might be part of the LLMs' training) as opposed to measuring their performance on previously unseen tasks. This approach improves the validity of the results and reveals the models' true limitations and capabilities. Finally, we compared the performance of six state-of-the-art LLMs, varying in parameter size, and reported their results across four distinct tasks using both quantitative and qualitative analyses.

\subsubsection{Structure of the Article:}
The rest of the paper is structured as follows: in Sec.~\ref{sec:relatedwork} the role of LLMs in Software Engineering (SE) tasks and required Verification and Validation process are discussed to identify research gaps. In Sec.~\ref{sec:methodology}, we outline the methodology used in this research followed by the results of the experiment in Sec.~\ref{sec:results}. Finally, we discuss the results in Sec.~\ref{sec:discussion} together with our proposed process to assure the safety of the LLM-generated code. We conclude our work in Sec.~\ref{sec:conlusion}.

\section{Related Work and Background} \label{sec:relatedwork}

\subsection{Large Language Models for Software Engineering} 

LLMs are already successfully applied to a wide range of SE tasks, including code comprehension and summarization~\cite{kumar2024code}. Hence, LLMs are considered as valuable assistants that provide support and insights before a human developer formulates the final artifacts~\cite{cabrero2024exploring} or specify the requirements~\cite{nouri2024engineering}.
LLMs can also augment the implementation of autonomous driving functions. For instance, Liu et al.~\cite{Liu2023EmpiricalStudy} explored LLM-based safety-related code generation for vehicle behaviour and concluded that LLMs can automatically generate safety-critical software codes.
LLMs have also been employed in automated vulnerability fixing~\cite{sagodi2024reality} and code repair~\cite{Liventsev2023} to efficiently fix software bugs without human intervention. However, manual verification comes with the cost of labour and the involvement of experts, typically in multiple iterations, as the LLM-generated code might not match the minimum expected quality~\cite{nouri2024engineering}.
Although LLMs can quickly generate code, we must not overlook their tendency to hallucinate, i.e., generating nonsensical or unfaithful information that does not align with the provided context or real-world knowledge~\cite{ashani2024llms}. 
As the adoption of AI grows, the need for AI to be trustworthy is also increasing~\cite{fernandez2021trustworthy}. Trustworthy AI can be defined as a conceptual framework that ensures that the development and implementation of technically and socially robust AI systems adhere to all relevant and applicable laws and regulations, and conform to general ethical principles~\cite{aihleg,aiact,ALKS}. These requirements apply not only for the integration of LLMs into software products and systems, but also the use of LLMs tools in software development~\cite{aiact}.

The capabilities of LLMs are often studied against existing benchmarks~\cite{siddiq2024using}, and then ranked using metrics such as pass@k~\cite{lyu2025top}, which employ predefined test cases and acceptance criteria.
However, as highlighted by Zhou et al., there are concerns about the benchmarks' data being used in LLM training~\cite{zhou2023don}, which is known as \textit{benchmark leakage}. This raises questions about LLMs' ability to handle unseen tasks, crucial for safety. Therefore, new evaluation methods as we propose in Sec.~\ref{sec:methodology} are needed to ensure that LLMs can meet expectations when employed for real world use-cases.

\subsection{Verification and Validation of Autonomous Driving Systems}

Verification and Validation (V\&V) are critical processes in the development lifecycle of Autonomous Driving Systems (ADS). Given the complexity and safety-related nature of ADS, it is imperative to ensure that these systems function correctly under a wide variety of real-world conditions. Effective V\&V guarantees that the system meets its intended requirements and performs as expected in diverse driving environments.
% Zoom in: Simulation and SIL
In autonomous driving, V\&V approaches can be categorized into multiple methods, each serving different purposes and suitable for different stages of development from unit verification to verification of software integration. These include, but are not limited to, code review, formal verification, and testing. In testing, the software is evaluated in different environments, such as:

\begin{itemize}
    \item Software-in-the-Loop (SIL): Where the software is integrated and tested within a simulated environment, without the need for physical hardware.
    \item Hardware-in-the-Loop (HIL): Involves testing with real hardware components integrated into a simulated system.
    \item Vehicle-in-the-Loop (VIL): Involves testing with an actual vehicle, typically in a controlled environment with real-world driving scenarios.
\end{itemize}

SIL facilitates and accelerates early Verification and Validation (V\&V) during the conceptual development of a product function, a stage traditionally addressed later in the development cycle, typically during the ``right leg'' of the V-model. By enabling rapid iteration and validation of software functionality within a controlled virtual environment, SIL allows for spotting and fixing potential issues long before hardware integration begins.

We use Environment Simulator Minimalistic (\textit{esmini}) \cite{website:EsminiGithub2018}, an open-source simulation tool designed for testing and validating advanced driver assistance systems (ADAS) and automated driving systems (ADS) during the SIL validation study. \textit{esmini} serves as a lightweight, flexible, and efficient platform for simulating complex traffic scenarios, making it particularly useful for Software-in-the-Loop (SIL) testing. \textit{esmini} is built on the OpenDRIVE and OpenSCENARIO standards, which are growingly used for defining road networks and dynamic traffic scenarios, respectively. This enables users to simulate real-world driving environments and interactions. In SIL applications, \textit{esmini} allows for virtual testing of ADAS/ADS functionalities by replicating various traffic situations, sensor inputs, and vehicle behaviours~\cite{fei2024SiLesmini}.
In order to apply SIL for the early verification of prototype code, its credibility should be ensured through rigorous validation methods. 
This alignment is crucial to ensuring that that the results of SIL-based validation can be trusted sufficiently, and they correlate with reality.
Correlation of selected SIL environment in this study is addressed in~\cite{fei2024SiLesmini} by proposing techniques for aligning high-fidelity, non-deterministic test track data with low-fidelity, deterministic simulations.

\section{Methodology} \label{sec:methodology}

\begin{figure*}
  \includegraphics[width=\textwidth]{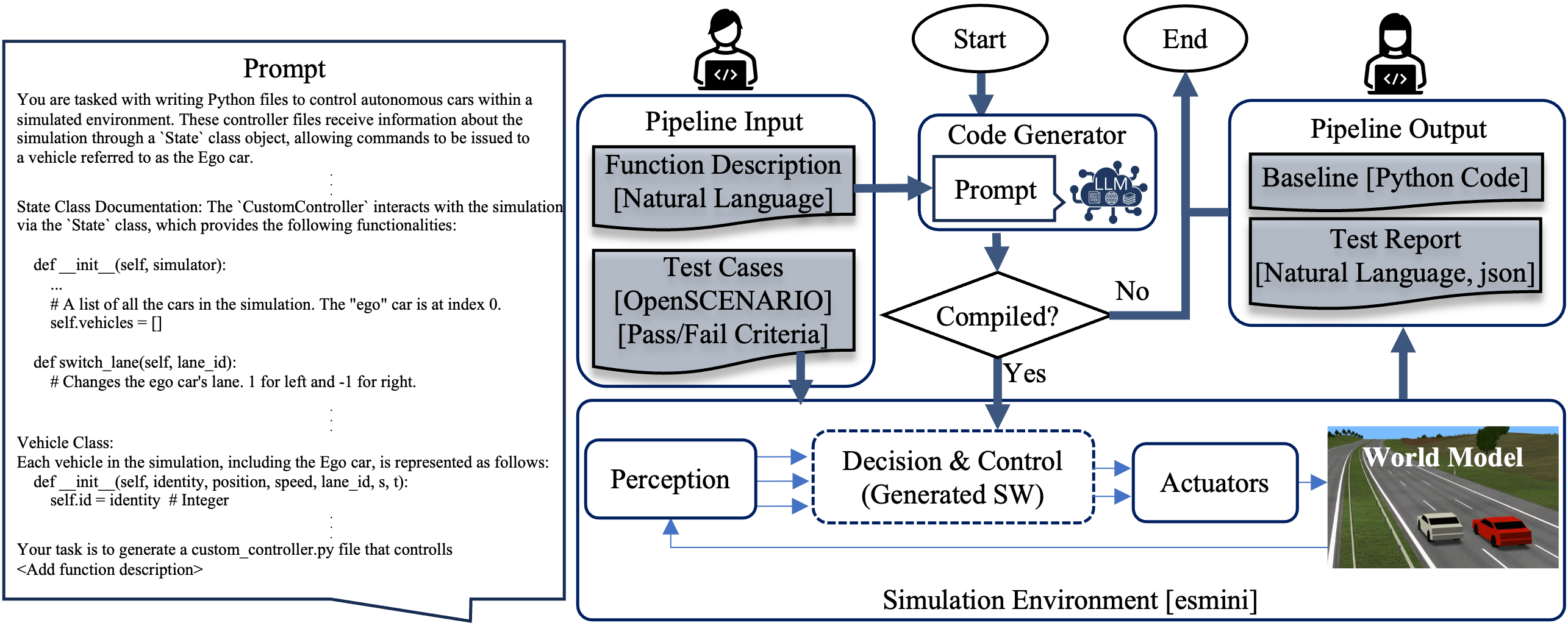}
  \vspace{-0.5cm}
  \caption{Designed and implemented pipeline, including the LLM model, prompts, and simulation environment. The function description is automatically inserted into the prompt, sent to the LLM, and the generated Python code is extracted from the LLMs' responses. Then, the compilable codes are sent to \textit{esmini} and tested against relevant test cases for the specific function. Finally, a report is generated and attached to the code as the output of the pipeline.}
  \label{fig:AbsVCodeSim}
\end{figure*}

\subsection{Technical Setup: Code Generation Pipeline}

As it is presented in Fig.~\ref{fig:AbsVCodeSim}, the proposed pipeline enable ``Evaluate fast, Eliminate fast'' concept for reaching the correct generated software faster as the remaining part of generation will be done by user and other testing environments. The implemented pipeline is configurable for different LLM architectures, enabling us to examine and compare the performance of multiple LLMs.

Five open-source LLM models (retrieved using Ollama) and GPT-4 (Azure-based via API) ~\footnote{deepseek-coder:33b (ID: acec7c0b0fd9); deepseek-r1:32b (ID: 38056bbcbb2d);  codegemma:latest (ID: 0c96700aaada); mistral:7b (ID: f974a74358d6); codellama:34b (ID: 685be00e1532); gpt-4 (version:turbo-2024-04-09)} were tasked with generating code for a controller for four automotive functions. The models were selected from a pool of the most well-known and frequently downloaded models on Hugging Face, following preliminary tests to identify the strongest candidates.
Four alternative prompts were used for each model to evaluate their performance and analyse their limitations. The most effective prompt was then selected, and all models were tested using \textit{esmini}. The evaluation was conducted according to the following process:

\begin{itemize}
    \item Does the generated controller code compile? 
    \item Does the generated controller code run? 
    \item Does the generated controller code pass all of the test cases? 
\end{itemize}

The acceptance criteria for all test cases are checked automatically in the logged data from each test case, and the success rates for compilation, execution, and passing test cases are stored in JSON format for each generated code. This quantitative evaluation is accompanied by an analysis of the responses, generated controller codes, and log files of the LLMs to better understand their limitations and drawbacks.
Additionally, the failed generated codes are analyzed to identify root causes, which are reported in the results section.

Since Python requires less code for the same task compared to lower-level programming languages (e.g., C++), it is a suitable choice for the designed function. This helps to reduce the risk of failure of experiment due to token limitations, especially when models with different number of parameters are being compared. Moreover, since Python is not suitable for the final software, using LLM-generated code in Python reduces the risk of code leakage (e.g., copy-pasting the code) into production-related software before gaining sufficient confidence in the technology and identifying its limitations and strengths.
Finally, Python demonstrates an average performance, with Copilot (GPT-based) achieving an output quality of  42\% compared to other languages( Java (57\%), JavaScript (27\%), and C (39\%)~\cite{Nguyen2022EmpiricalEvalCopilot}). 
This makes Python a strong candidate for evaluating the average performance of LLMs, which contributes to the generalizability of the findings.

\subsubsection{SIL with \textit{esmini}:}

As \textit{esmini} supports a fast verification of system requirements and functionality using prototype code, it comes naturally as preferred choice for also testing LLM-generated code. Combining LLM-based code generation with SIL in a feedback loop ensures that functional components are validated early, reducing the time and costs associated with hardware-based testing while streamlining the transition to later stages of development.
It is important to note that the focus of early verification of prototype code is not on achieving the full test coverage, but rather intended to deliver the generated code to human for code review. 
The selected simulation environment allows the engineers to also visualise the behaviour of the function. This way, it enables them to check whether the intentions of the function designer are met. Furthermore, the simulation environment also let the engineer to test the safety goals or functional safety requirements.

\subsection{Code Generation Experiments: Tasks and Test Cases}

To ensure the successful integration of LLM-generated code into an existing software system, the LLM must accurately adhere to the defined interface requirements. Moreover, the generate code shall calculate and request the expected output signals with the correct timing and precision.
Hence, by analysing the results of our preliminary experiments on code generation tasks, we identified nine key capabilities required for successful code generation, as listed in Table~\ref{MapCapabilities&Functions}. Accordingly, four functions are specified with varying levels of difficulty and complexity to be implemented by the LLM. As mapped in Table~\ref{MapCapabilities&Functions}, these functions allowing us to evaluate the capabilities necessary for LLMs in a real automotive application.
The functions are specified as follows:

\begin{itemize} [leftmargin=0.75cm]
    \item [\textbf{F1}:] The ego vehicle shall start braking if the speed exceeds 10 m/s2.
    \item [\textbf{F2}:] The ego vehicle shall perform a lane change to the right if there is a vehicle in the same lane as the ego vehicle.
    \item [\textbf{F3:}] The ego vehicle shall adapt its speed to the vehicle in front to avoid collision (exhibiting so called Adaptive Cruise Control behaviour, ACC).
    \item [\textbf{F4}:] In unsupervised Collision Avoidance by Evasive Manoeuvre (CAEM), the ego vehicle shall perform a lane change to avoid imminent collision with the vehicle in front. The lane change is preferably to be conducted to the left.
\end{itemize}

As presented in Table~\ref{MapCapabilities&Functions}, F4 is designed as the most complex function, covering all identified capabilities. Moreover, since unsupervised lane change (i.e., decision-making for when to change lanes) is an advanced feature, the risk of LLMs being trained on such a dataset is minimal.

\begin{table}[h]
    \centering
    \caption{Presents the list of identified capabilities ($C_x$) required to generate a code with an acceptable level of maturity to be delivered to an engineer for code review. These functions ($F_x$) are selected or designed to test a subset or all of these capabilities at different levels of complexity. F1 and F2 are designed to capture the minimum capabilities of less capable LLMs. While F3 and F4 are more complex and examine the LLMs' performance in real industrial use cases.}
    \label{MapCapabilities&Functions}
    \resizebox{\textwidth}{!}{ % Resizes table to fit within page width
    \setlength{\tabcolsep}{3pt}
    \begin{tabular}{|l|p{8.5cm}|c|c|c|c|} % Custom column widths
        \hline
        \textbf{$C_x$} & \textbf{Capability Description} & \textbf{F1} & \textbf{F2} & \textbf{F3} & \textbf{F4} \\
        \hline
        C1 &Input: Reading ego state (e.g., speed) & X & X & X & X\\
        \hline
        C2 &Logic: Decision based on ego state & X & X & X & X \\
        \hline
        C3 &Output: Calculation of continuous request (e.g., speed, lane change) or deciding on a binary output (e.g., brake) & X & X & X & X \\
        \hline
        C4 &Input: Reading other agents' states (e.g., speed or position) & - & X & X & X \\
        \hline
        C5 &Logic: Decision based on ego and other agents' states  & - & X & X & X \\
        \hline
        C6 &Logic: Precise calculation of request based on current state of ego or others & - & X & - & X \\ %lateral acceleration is weird (check figures)
        \hline
        C7 &Closed loop adaptation: Adapting the request based on the action of agents (speed control, or double lane change) & - & - & X & X \\
        \hline
        C8 &Logic: Need to follow traffic rules (overtake from left) & - & - & - & X\\
        \hline
        C9 &Logic: Prioritising  safety over traffic rules if needed (e.g., overtaking from right in case a safety requirement will be violated) & - & - & - & X \\
        \hline
    \end{tabular}
    }
\end{table}

\subsection*{Requirements for the Generated Code}

The generated code shall satisfy the following requirements ($R_x$):

\begin{itemize} [leftmargin=0.75cm]
    \item [\textbf{R1}:] The generated code shall be integrable in a predefined software architecture without any manual modification or improvement (e.g., APIs and outputs).
    \item [\textbf{R2:}] The generated code shall control only the specified signal in the described function and must not affect any other signals.
    \item [\textbf{R3}:] The generated code shall not request any actions that could lead to exiting the drivable area (e.g., going off the road).
    \item [\textbf{R4:}] The generated code shall decide on proper action to avoid collisions with other static or dynamic objects (intended only for ACC and CAEM).
\end{itemize}

\subsection*{Test Cases} \label{sec:tests}

Considering the potential malfunctions of CAEM and ACC, seven scenarios have been designed to test the generated codes, each covering multiple requirements:

\begin{itemize} [leftmargin=1.4cm]
    \item [\textbf{TC1-3}:] Three multi-lane highway scenarios, with the ego vehicle driving at 120 kph (TC1), 80 kph (TC2), and 40 kph (TC3). A second vehicle performs a cut-in manoeuvre and immediately decelerates. The time to brake in the scenarios is 0.4 seconds, shorter than the time-to-react by a skilled driver according to UNECE Regulation No.~157~\cite{ALKS}.
    \item [\textbf{TC4-5}:] Similar to TC1-3, a second vehicle overtakes the ego vehicle and brakes in front of it. However, in TC4 and TC5, a third vehicle (static or matching the ego-vehicle speed, respectively) is blocking the ego vehicle's avoidance manoeuvre.
    \item [\textbf{TC6-7}:] In TC6, there are no other vehicles on the road, while in TC7, a second vehicle is in a parallel road, driving in the opposite direction.
\end{itemize}

\textbf{TC1-3} are intended to test the ``commission'' malfunction of deceleration (in ACC) and lane change (in CAEM). \textbf{TC4-5} are designed only for CAEM to test the capability of the generated code in handling more complex scenarios with multiple agents and covering malfunctions such as ``delayed'' or ``wrong lane change.''
To prevent the LLM from generating code tuned to the test cases (test case leakage), the description of these tests cases are not included in the prompts. Finally, \textbf{TC6-7} are intended to test ``commission'' malfunctions, which include unintended lane changes and deceleration for ACC and CAEM, respectively~\footnote{Appendix providing the detailed and visual descriptions of test cases, and sample generated codes for all configurations: \url{https://doi.org/10.5281/zenodo.14913411}}

\subsection*{Experimental Setup}
Six configurations of the pipeline are evaluated, each corresponding to one of the models described earlier. Each configuration is independently employed to generate code for one of the four automotive functions defined in Section 3.2, one at a time. The description of the specific function in the setup is added at the end of the prompt shown on the left side of Fig.~\ref{fig:AbsVCodeSim}.
This combination results in 24 unique experimental setups (6 models × 4 functions).
To mitigate the inherent stochastic nature of the LLM output, each set-up was executed 20 times. An experiment run is considered successful only if the generated code passes all test cases without violating any of the predefined requirements (R1–R4).

\section{Results} \label{sec:results}

We first analyse the performance of the six state-of-the-art LLMs on F1 and F2. As shown in Table~\ref{MapCapabilities&Functions}, F1 and F2 are designed to assess the minimum capability of the models.

\begin{figure*}
  \includegraphics[width=\textwidth]{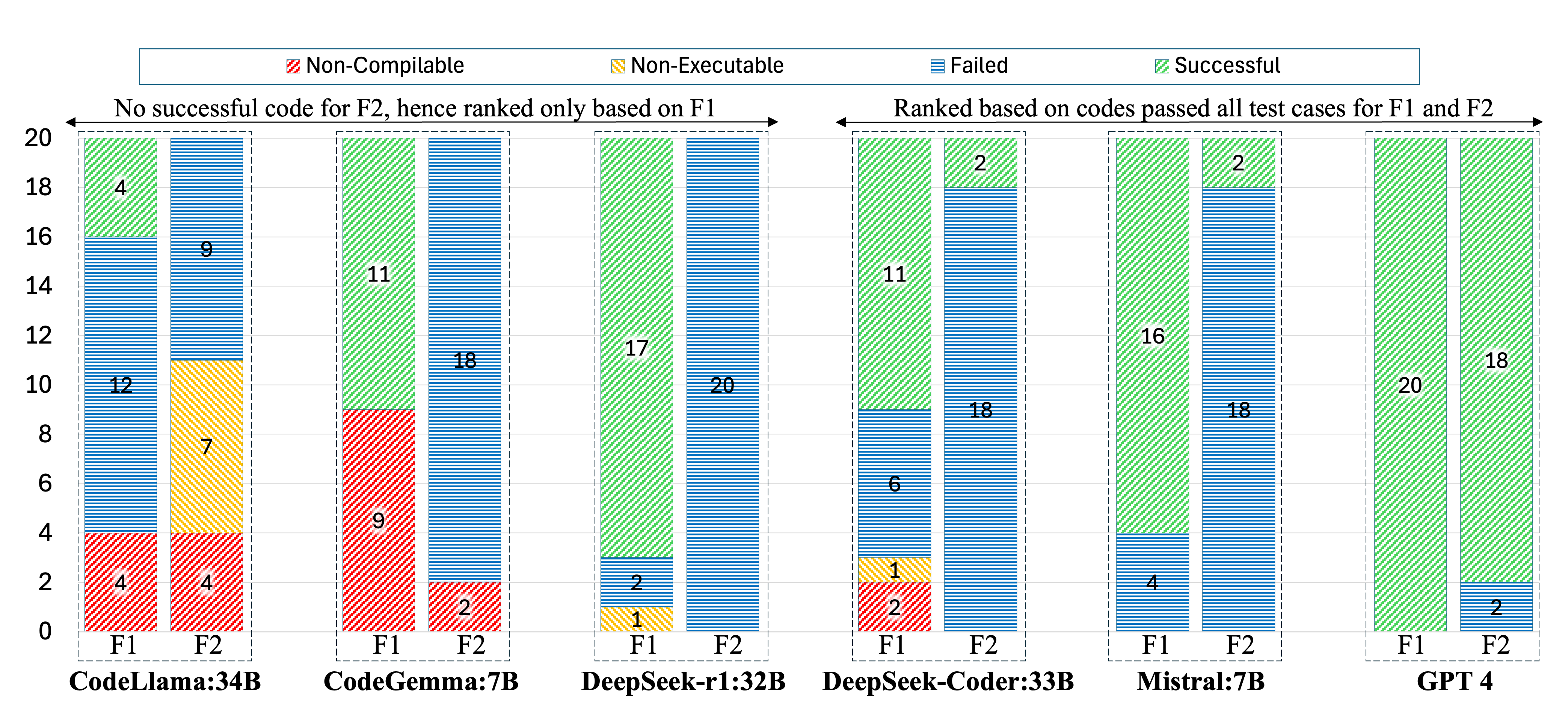}
  \caption{Reports the performance of all LLM models on two simple functions (F1 and F2). The left bar of each model presents the results for F1 (i.e., brake if the speed is higher than 10 m/s²), and the right bar presents the results for F2 (i.e., lane change until reaching the rightmost lane). The performance of the models is ranked first based on the total number of successful codes for F2 and then based on F1, as F2 is considered more complex than F1.}
  \label{fig:ResultsF1F2}
\end{figure*}

For F1, as reported in Fig.~\ref{fig:ResultsF1F2}, GPT-4 delivered 20 fully successful codes for F1 and 18 for F2.
All other models generated between 4 (CodeLlama:34B) and 17 (DeepSeek-r1:32B) successful codes (green area). This demonstrates their varying ability to generate code that reads relevant inputs and triggers appropriate outputs based on the ego vehicle's state. It confirms that the models are capable of interpreting the given inputs and outputs as described in the prompt. Moreover, some models exhibited creative solutions despite limited flexibility in the given task. For instance, Mistral:7B generated two codes that reversed speed instead of braking, while DeepSeek-Coder:33B (in one case) and DeepSeek-r1:32B (in two cases) reduced speed to avoid a collision rather than applying the brakes. However, since braking was explicitly required, these solutions were classified as failures due to non-compliance requirement R2.

For F2, DeepSeek-Coder:33B and Mistral:7B were capable of delivering successful code (2 cases), while the other open-source LLM models failed to generate any successful code. The complexity of F2 increased compared to F1, as it required additional capabilities such as reading the state of other agents through inputs, understanding the position of relevant agents (e.g., detecting the presence of other agents in the ego's lane), and deciding on appropriate actions (e.g., changing the lane to the right). Furthermore, F2 demanded precise lane-change values, as lateral motion is more sensitive than longitudinal adjustments. For instance, some failed cases occurred due to excessive lane changes. If the code instructed a single lane change in two steps, it resulted in an unintended two-lane change, causing the vehicle to exit the drivable area.
Thus, we observe the following potential risks from LLM-generated code:
\begin{itemize}
    \item Alternative strategies (e.g., reducing or reversing the speed) instead of what was explicitly asked for (e.g., braking).
    \item Failing to retrieve the state of the ego vehicle from the available interfaces.
    \item Unnecessary (e.g., double lane change) or illegal (e.g., leaving the drivable area) manoeuvres that conflict with the requirements.
\end{itemize}

\begin{figure*}
  \includegraphics[width=\textwidth]{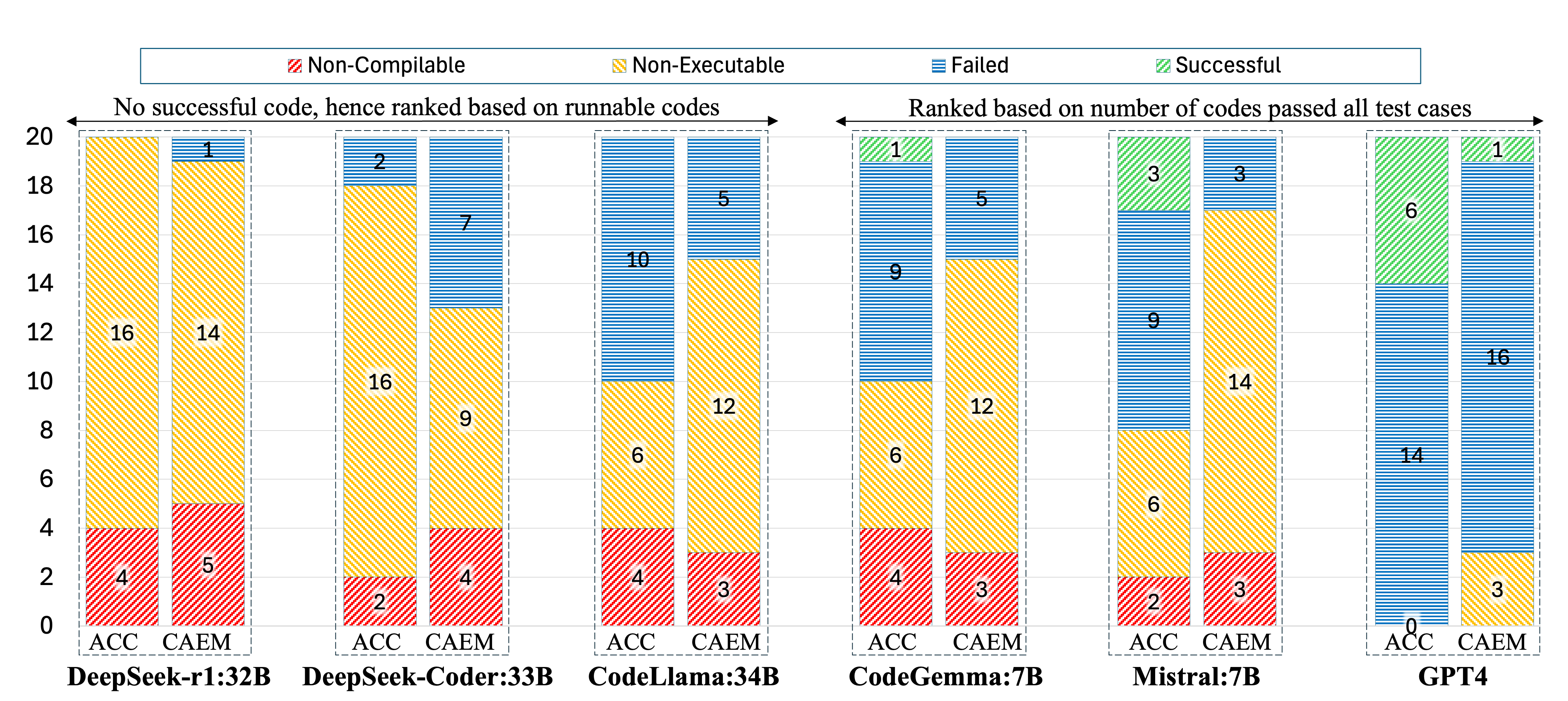}
  \caption{Performance of six models on two advanced automotive functions (F3 and F4). The ACC bar in each group indicates the models' performance in F3, and the CAEM bars present the results for F4. The models are ranked first based on total successful generations for F3 and F4, and then by the number of executable code generations.}
  \label{fig:ResultsACC_CAEM}
\end{figure*}

In the second experiment, all models are examined by more advanced automotive functions: F3, adapting the speed to the vehicle in front (ACC), and F4, lane change to avoid imminent collisions (CAEM). Several models successfully generated code for F3 (ACC): Mistral:7B and CodeGemma:7B respectively generated 3 and 1 successful code versions, while GPT-4 was able to generate 6 successful codes. As reported in Fig.~\ref{fig:ResultsACC_CAEM}, the only LLM, which could deliver the code for F4 (CAEM), was GPT-4, while all others failed.

Contrary to the open-source models, only 3 of the code versions generated by GPT-4 for F3 and F4 contained errors that led to non-executable code, while the rest did not pass the tests described in Sec.~\ref{sec:tests} (blue area). Some of these failures were due to incorrect values selected for thresholds, such as the safe distance threshold or time-to-collision. This led to late lane changes, which could be fixed by changing the values. These values require tuning or must be derived from standards, regulations, or expert domain knowledge within the company. We did not provide the models with any hints on these numbers, as doing so could bias them toward a specific solution. Hence, the 5\% success rate of GPT-4 for CAEM could be improved by adjusting the thresholds. However, this was not done to avoid compromising the validity of the experiment. 
Comparing the reported results of the models for F1 and F2, in Fig.~\ref{fig:ResultsACC_CAEM}, with the results for ACC and CAEM, in Fig.~\ref{sec:results}, we note the impact of the task complexity on a model's performance. Both DeepSeek models dropped in ranking from second and third place to last for ACC and CAEM. Moreover, higher task complexity does not only increases the failure rate of test cases but also affects the code quality, reducing the likelihood of generating executable or compilable code.

The failed codes were also analysed to identify and report the most common root causes of failures. Most failures in the open-source models were due to syntax errors or incorrect output calculations, leading to failed compilation, execution errors, or no action in the simulation. For instance, in some cases, CodeLlama-34B and Mistral did not generate any code in the response, but instead returned a natural language explanation of the logic or just a skeleton of the code with comments. Using incorrect syntax to access attributes or methods (e.g., $ego\_car.position[0]$ instead of $ego\_car.s$) are also observed in the generated codes, even when explicitly provided in the prompt.
Another cause of compilation or execution failures in the simulation was the addition of unnecessary extra code that prevented integration. For instance, in six occasions, DeepSeek-r1:32B hardcoded the test case within the generated code, causing conflicts with the simulation. Additionally, most models failed to account for an edge case where the relative speed between the ego vehicle and the vehicle in front could be zero. As a result, when calculating the time to collision, they attempted to divide by zero, leading to a division-by-zero error and a subsequent simulation execution failure. Other code failures were due to incorrect logic, e.g., CodeGemma often incorrectly assumed that the first car in the object list was the vehicle in front, without considering its longitudinal and lateral position.
We observe the following potential risks for safety-critical, automotive functions:
\begin{itemize}
    \item Choosing incorrect threshold values (e.g., safety distance between cars) within the code that might lead to danger (e.g., late lane changes).
    \item Task complexity and originality might negatively impact the performance of otherwise well-performing models.
    \item Failing to use the specific syntax needed to access the provided interfaces.
    \item Generating unnecessary code that prevents integration with the simulation tool or provided interface.
    \item Failing to consider edge-cases that lead to run-time errors (e.g., dividing by the relative speed when it is zero).
\end{itemize}

\section{Discussion} \label{sec:discussion}

In F2 and CAEM, the code shall request the exact value (1 for left and -1 for right) for lane changes ($state.switch\_lane(x)$) within the required time. Additionally, the requested lane change must not cause the ego vehicle to exit the drivable area (lane ID: -2 to -4), This means that a single extra lane change request could lead the ego vehicle off the road.
Hence, the sensitivity of F2 and CAEM are high to the requested output as mentioned in Table~\ref{MapCapabilities&Functions}. On the contrary for F1 and ACC, if the generated code requests for a slightly wrong value, the function might survive and this might be one of the root causes for the high success rate of F1 and ACC. 
As discussed, the risk of benchmark leakage justify the design of CAEM as a complex function, which requires all capabilities in Table~\ref{MapCapabilities&Functions}, and it is also not as publicly available and known as ACC.
We in general observe that LLMs seem to be not yet fully ready to deliver code for function development, however they show promising potential. For instance, GPT-4 delivered one successful code out of 20 attempts for the CAEM function, while others were not successful at all. 
As presented in Sec.~\ref{sec:results}, the ranking of the models based on their performance for simple functions changes when compared to more complex functions. Moreover, it seems that the number of parameters has less impact on the performance of the models.

\begin{figure*}
  \includegraphics[width=\textwidth]{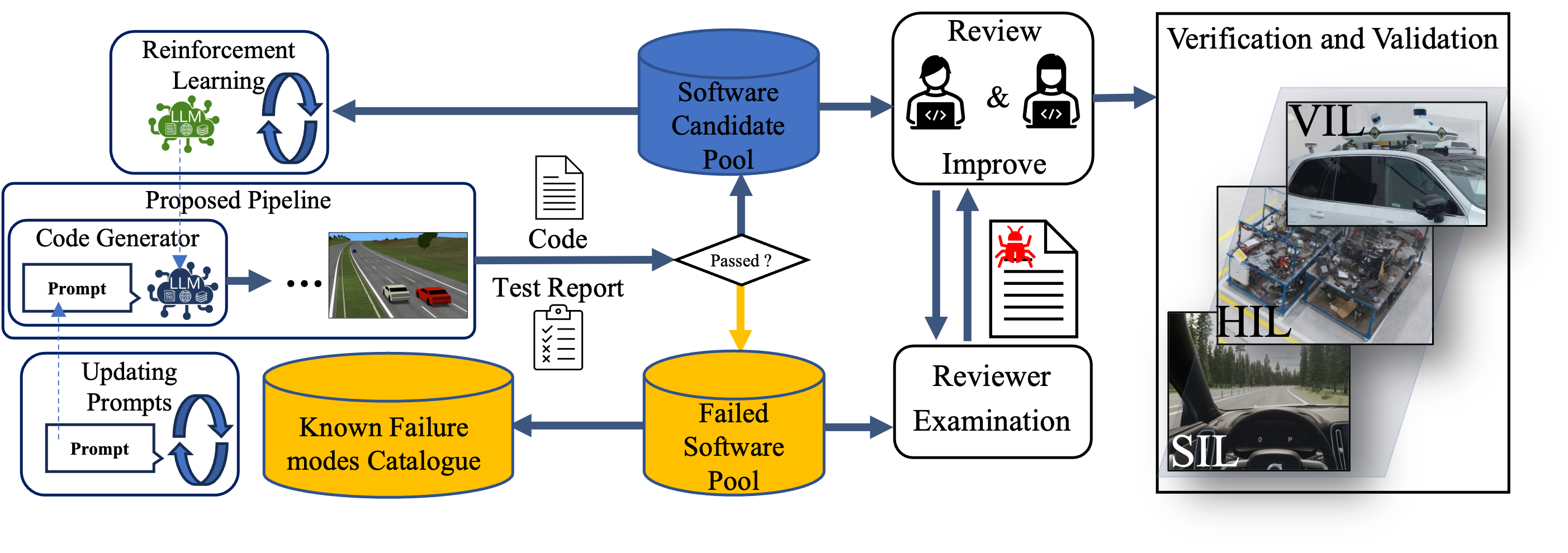}
  \vspace{-0.8cm}
  \caption{Pipeline in Fig.~\ref{fig:AbsVCodeSim} integrated into the software engineering process. It enables the pre-evaluation of generated codes, with the best candidates selected for engineers to review and improve before proceeding to the rigorous V\&V process (right side).
Failed codes are analysed to extract failure modes, helping refine the prompts. To avoid automation bias and evaluate the effectiveness of the review process, the failed codes can be provided to check whether they are detected and excluded.} %Bea I am talking with Chritisan right now: to check how effective is human oversight for EU AI act.
  \label{fig:VRandCR}
\end{figure*}

\subsubsection{LLM-augmented Review Process:} \label{sec:VRandCR}
Currently, LLMs are directly used by the software developers, and might generate non-compilable, non-executable, or even unsafe code (i.e., seemingly working, but violating safety requirements).
In order to benefit from the proposed pipeline, we suggest a process as illustrated in Fig.~\ref{fig:VRandCR} to safeguard the integration of LLM-generated code into the development of automotive functions.  
Asking human reviewers to assess any LLM-generated code, which is usually a complex and lengthy task, may become a waste of time, especially since, as seen in Fig.~\ref{fig:ResultsF1F2} and~\ref{fig:ResultsACC_CAEM}, most of the generated versions of the code are not directly usable. Hence, the code shall be pre-evaluated by a set of simple, fast, and yet relevant, verification tools such as \textit{esmini}. Then, the best candidates are sent to engineers for review and improvement. Human-oversight is a crucial part of this process even if the proposed pipeline can generate fully functional code, as also recommended by the EU AI Act~\cite{aiact}.
The preliminary evaluation of the generated code versions, as seen in Sec.~\ref{sec:results}, can also help to extract a failure modes catalogue, such as the one extracted from the two experiments above.
Such a failure modes catalogue could also help refine prompts and highlights the need for tools to detect bugs and to mitigate recurring failures. 
The ranked codes would be used in improving the LLMs themselves through reinforcement learning.
This process may help LLMs and humans in generating the final version of the code, in which low-ranking code versions can also be part of the human review process in order to fight automation bias (so humans do not become too complacent with LLMs generating more and more parts of the final code base).

\subsubsection{Threats to Validity:}
To ensure the \emph{construct validity} of the employed evaluation method, the selected simulation environment (\textit{esmini}) is part of the correlated SIL toolchain used in industrial AD development, as reported in~\cite{fei2024SiLesmini}.
The specific tasks (F1-F4) are designed to reduce the risk of data and benchmark leakage as much as possible to evaluate the capabilities of the LLMs in handling new tasks, which is contributing to \emph{internal validity}. 
Publicly available elements of the pipeline, such as OpenSCENARIO and \textit{esmini}, would not constitute leakage, since the model is not asked to reproduce the scenario or simulation environment, but rather to generate code for functions. Importantly, if an LLM is capable of leveraging its understanding of the simulation environment or scenario structure to produce valid control logic, this indicates reasoning and generalization capability, which is an advantage. Benchmark leakage only occurs when models reproduce memorised solutions rather than generate new code from learned understanding.
Moreover, we conducted additional tests to identify potential threats to validity in the experiment such as removing the interface descriptions from the prompt to examine whether the LLMs had encountered similar code in the esmini environment.
As demonstrated by the 24 experiment configurations (6 LLM models generating code for 4 functions), the pipeline is adaptable for model- or function-agnostic use. Thus, it can be applied to other domains, programming languages, and verification environments, addressing \emph{External Validity}. However, since model performance varies by task, each configuration must be evaluated separately.

\section{Conclusion} \label{sec:conlusion}

Examining the performance of LLMs against known benchmarks is a common method; however, the risk of these models being trained on publicly available data is high, which is a phenomenon known as benchmark leakage. This could lead to false expectations when using them to generate code meant for production. Therefore, it is crucial to establish more reliable processes for testing the outcome of these tools to better capture their strengths and weaknesses.
To address this, we proposed an LLM-based pipeline that includes a simulation model for iterative, seamless, and automatic evaluation of the generated code.
The pipeline enables ``evaluate fast, eliminate fast,'' resulting in multiple pre-evaluated candidates before being sent for code review to meet the human oversight as demanded by the EU AI Act.
We designed four functions with varying levels of complexity to test different basic capabilities required for programming and reported on the results.
As open-source models were not successful with the most complex function, they are examined for simpler functions to explore their minimum capabilities to generate functional code for safety-critical, automotive functions. Hence, F1 and F2 were designed to assess capabilities such as reading the ego vehicle's state and acting accordingly.
We finally evaluated the performance of LLMs in generating code for both longitudinal (F3) and lateral (F4) decision \& control of an autonomous vehicle. 
As reported in Fig.~\ref{fig:ResultsACC_CAEM}, only GPT-4 was capable of generating successful code that passed all test cases for a complex function, while other models failed even to generate code that just compiles.
However, the evaluation results for F1 and F2, presented in Fig.~\ref{fig:ResultsF1F2}, indicate that open-source models can generate successful codes for simpler functions and we expect further improvements in the near future anticipating the current rate of innovation in this area.
As demonstrated by these experiments, there is a need for trade-off between the complexity of required tasks and the selected LLM. In some cases, a model with seven billion parameters (e.g., Mistral:7B) is sufficient, while for higher complexities, even a model with orders of magnitude more parameters (GPT-4) still struggles to generate code.

As expected, due to the stochastic nature of LLMs, the generated code fails to produce outputs that pass all test cases and satisfy the defined requirements (R1–R4) in 15\% to 100\% of the cases. As the experiment was conducted across 24 setups, the success rate varies depending on the complexity of the task and the model's capability. A high success rate in one case is not generalisable to other cases, as training data and task complexity are the influencing factors. Based on our findings, we advise to reduce overly relying on such LLM-based code generation pipelines for the moment; however, again anticipating the rate of innovation in this area, in combination with an iterative, fully automated self-assessment cycle using virtual testing methods, we expect tremendous impact of the role how automatic code generation will play in future development processes--even for safety-critical functions.

\begin{credits}
\subsubsection{\ackname} This work has been partially supported by Sweden’s Innovation Agency (Vinnova, diarienummer: 2021-02585), and by the Wallenberg AI Autonomous Systems and Software Program funded by the Knut and Alice Wallenberg Foundation.

\subsubsection{Disclaimer}
The views and opinions expressed are those of the authors and do not necessarily reflect the official policy or position of Volvo Cars.
\end{credits}

\end{document}